\documentclass[aps,prl,showpacs,superscriptaddress,groupedaddress]{revtex4}  
\usepackage{graphicx}  
\usepackage{dcolumn}   
\usepackage{bm}        
\usepackage{amssymb}   
\usepackage{amsmath}
\usepackage{makeidx}
\usepackage{amsfonts}
\usepackage{amssymb}
\usepackage{dsfont}
\usepackage{color,soul}
\usepackage[autostyle]{csquotes}
\begin{document}
\title{Role of the Interplay Between the Internal and External Conditions in Invasive Behavior of Tumors}
\author{Youness Azimzade}
\affiliation{Department of Physics, University of Tehran, Tehran 14395-547, Iran}
 \author{Abbas Ali Saberi}
\email[Corresponding Author:~]{ab.saberi@ut.ac.ir} 
\affiliation{Department of Physics, University of Tehran, Tehran 14395-547, Iran}
 \affiliation{Institut f\"ur Theoretische Physik, Universitat zu K\"oln, Z\"ulpicher Strasse 77, 50937 K\"oln, 
 	Germany}
 \author{Muhammad Sahimi}
 \affiliation{Mork Family Department of Chemical Engineering Materials Science, University of Southern 
 	California, Los Angeles, California 90089-1211, USA}
\date{\today}
\begin{abstract}
Tumor growth, which plays a central role in cancer evolution, depends on both the internal features of 
the cells, such as their  ability for unlimited duplication, and the external 
conditions, e.g., supply of nutrients, as well as the dynamic interactions between the two. A stem cell 
theory of cancer has recently been developed that suggests the existence of a subpopulation of 
self-renewing tumor cells which is responsible for tumorigenesis, and is able to initiate metastatic 
spreading. The question of abundance of the cancer stem cells (CSCs) and its relation to tumor malignancy
has, however, remained an unsolved problem and has been a subject of recent debates. In this paper we propose a novel model beyond the standard stochastic models of tumor 
development, in order to explore the effect of the  density of  the CSCs and oxygen on
the  tumor's invasive behavior. The model identifies natural selection as the underlying
process for complex morphology of tumors, which has been observed experimentally, and indicates that 
their invasive behavior depends on {\it both} the number of the CSCs and the oxygen density in the
microenvironment. The interplay between the external and internal conditions may pave the way for a new 
cancer therapy. 
\end{abstract}
\maketitle

\section{Introduction}

Cancer usually begins with out-of-order duplication of a single cell that has stem cell-like behavior, referred to as the cancer stem cell (CSC)~\cite{reya2001stem}. Based on the CSC hypothesis, a CSC can duplicate unlimitedly and differentiate\cite{beck2013unravelling}. The classical CSC hypothesis proposes that, among all cancerous cells, only ``a few'' act as stem cells, but studies have reported\cite{quintana2008efficient} that a relatively high proportion of the cells were tumorigenic, contradicting the general belief. The CSCs have been proposed as the driving force for tumorigenesis and the seeds for metastases \cite{medema2013cancer}.  Their decisive role in maintaining capacity for malignant proliferation, invasion, metastasis, and tumor recurrence has been reported frequently\cite{li2014cancer}. For example, CSCs of breast tumor are involved in spontaneous metastases in mouse models\cite{liu2010cancer}. Also, CSCs promote the metastatic and invasive ability of melanoma\cite{lin2016notch4+} and their prsence is correlated with invasive behavior at colorectal adenocarcinoma\cite{choi2009cancer}. The effect of number of CSCs on tumor morphology has been subject to some experimental studies and simulations. Based on simulations\cite{enderling2013cancer,sottoriva2010cancer}, the frequency of the CSCs  smooths the morphology of tumor and based on experimental study\cite{castellon2012molecular}, the number of CSCs is higher in tumors with medium invasiveness (Gleason grade) than tumors with lower (Gleason grade) and higher (Gleason grade) invasiveness.  However, the relation between tumor malignancy and the frequency of the CSCs needs more  clarification\cite{medema2013cancer}.

Cancerous cells use  oxygen  to produce metabolites for duplication and growth\cite{vaupel1989blood}. Experimental \textit{in-vivo}\cite{hockel1996association} and \textit{in-vitro}\cite{cristini2005morphologic} studies, as well as computer simulations\cite{anderson2006tumor, anderson2005hybrid}, have reported that the density of oxygen regulates tumor 
morphology  and its shortage drives morphological irregularities. Due to the apparent strong correlations between the tumors' shape and their malignancy, fractal characterization of tumors has been used as a diagnostic assay for various types of tumors\cite{bru1998super, caldwell1990characterisation, lee2005predictive}. However, there is still no explanation as to why cellular structures at the scale of tumors display self-similar characteristics\cite{baish2000fractals}. 

In this paper we propose a novel model to study the effect of the number of the CSCs and the 
oxygen's  density on the invasive behavior of    general type of cancer. As we show below,  the development of  irregular shapes and respectively tumor's invasive behavior is correlated with these two factors.   
Unlike the previous studies, we present a quantitative measure by which one understands better the effect
of completion on the malignancy of tumors. We take the shape irregularity as the 
factor for identifying the invasive behavior of  tumor and compare our results with experimental
reports. The model that we present contains the essential
features of the cells, such as  symmetric/asymmetric division, metabolic state, cellular quiescence and 
movements, apoptosis, and  existence of oxygen and its consumption. Our results explain,
for the first time to our knowledge, the aforementioned experimentally-observed fractal behavior and 
contradict the predictions of recent models for the relation between the number of the CSCs and the 
growth rate and invasion. In addition, we believe that the results may cast doubt on the recent 
therapeutic approach based on oxygen deprivation. 

\section*{Results}  
\begin{figure}[h]
	\centerline{\includegraphics[width=1.\textwidth]{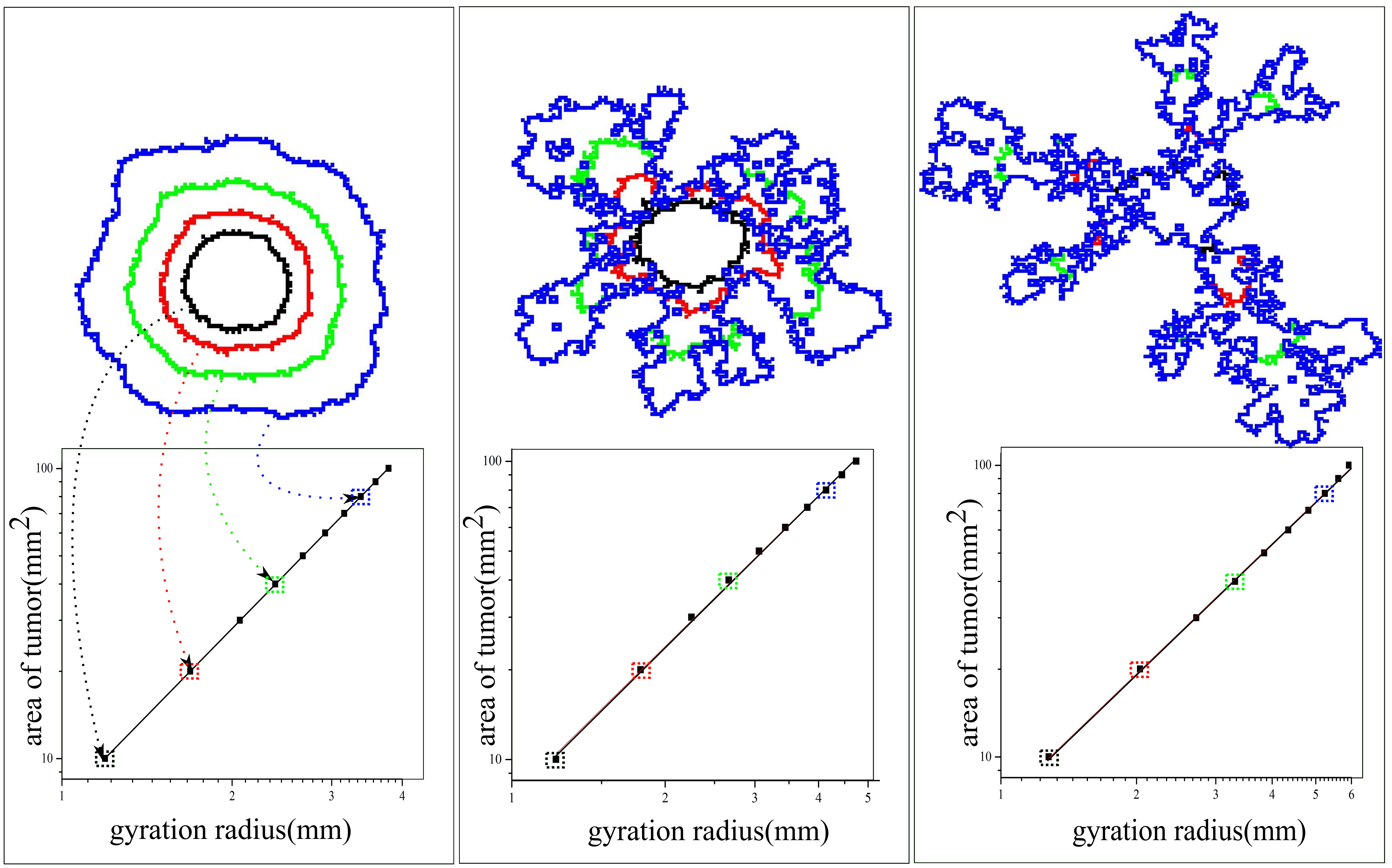}} 
	\caption{\textbf{Fractal structure of the tumors.} Tumors are irregular, but exhibit self-similarity. 
		The linearity of the plot indicates fractal behavior, with the slope being $D_f\approx 1.99\pm 0.01$ for 
		$p_s=0.1$ (left), $1.76\pm 0.02$ for $p_s=0.5$ (middle), and $1.47\pm 0.02$ for $p_s=1$ (right), with
		(normalized) oxygen density, $n=1$. Each contour line represents the borderline of the tumor 
		with the  corresponding gyration radius indicated by the dotted arrows. It should be noted that the left one covers 5000 units (50 mm$^{2}$) in 5000 time steps (30 days) while the middle one and the right one cover
			the same area in 14000(55 days) and 12000 (83 days) steps respectively. These simulations has been done within a 200 $\times $ 200 lattice}
	\label{FIG1}					
\end{figure}

As the system evolves, the cells consume oxygen, enhance their metabolic state, and proliferate after 
reaching the energy level of $u_p$, in order to create a clone ---the tumor--- see Figure\ref{FIG1}. 
The perimeter of this clone is the main object that we study in this paper. 

As Figure\ref{FIG1} demonstrates,  the cells take on irregular shapes during their growth whose complexity depends on the number of the CSCs ($p_s$).  One interesting approach is to study the structure of the borders in the context of interface instability \cite{vasiev2004classification, vasiev1994simulation, vasieva1994model}. The analogy with the instability of interfaces has been established for the case of melanoma\cite{amar2011contour} and the instabilities were attributed to nutrient density. But here, we are going to quantify tumor behavior through classifying irregular morphology of tumors. To quantify the irregularity of the tumor's morphology and its evolution, we use fractal analysis. To this end,  we measure the average distance $r$ from the center of the mass, as well as the area of the tumor during its growth.
Figure\ref{FIG1} indicates that $\log({\rm area})$ versus $\log r$ is a linear plot so that, ${\rm area}\sim r^{D_f}$. Thus, the slope of the line in the logarithmic plot is the fractal dimension $D_f$, implying self-similarity of the tumors of various sizes. The self-similarity of the tumors' growth is the result of heterogeneous duplication on their perimeter, which itself is due to the oxygen  gradient. Cells in the region with higher curvatures have better supply of oxygen, helping them increase their metabolic state and proliferate faster. The proliferation also creates new perimeter curvatures with the same behavior. As the number of oxygen  
consumers, which is proportional to $p_s$, increases the competition between the cells for the limited 
oxygen supply intensifies and  oxygen  availability becomes more heterogeneous. Thus, the
tumors take on more irregular shapes or lower fractal dimension $D_f$, contradicting the previous studies
\cite{sottoriva2010cancer, enderling2013cancer} that proposed an adverse relation between the number of 
the CSCs and the invasive behavior. 

We note that fractal scaling has been reported previously in the experimental studies 
\cite{caldwell1990characterisation, bru1998super}. Moreover, irregular shapes have been interpreted as 
an indication of invasive behavior of different tumors\cite{bru1998super, caldwell1990characterisation, 
	lee2005predictive}. Tumors with more irregular shapes are more invasive, and in our model the more irregular tumors have smaller $D_f$.  
There are several reports that confirm the correlation between $D_f$ and tumor malignancy (a malignant 
tumor possesses a lower fractal dimension than that of a benign mass) \cite{tambasco2010morphologic, etehad2010analysis, zook2005statistical, perez2015relationship}. 

\begin{figure}[h]
	\centerline{\includegraphics[width=0.001\textwidth]{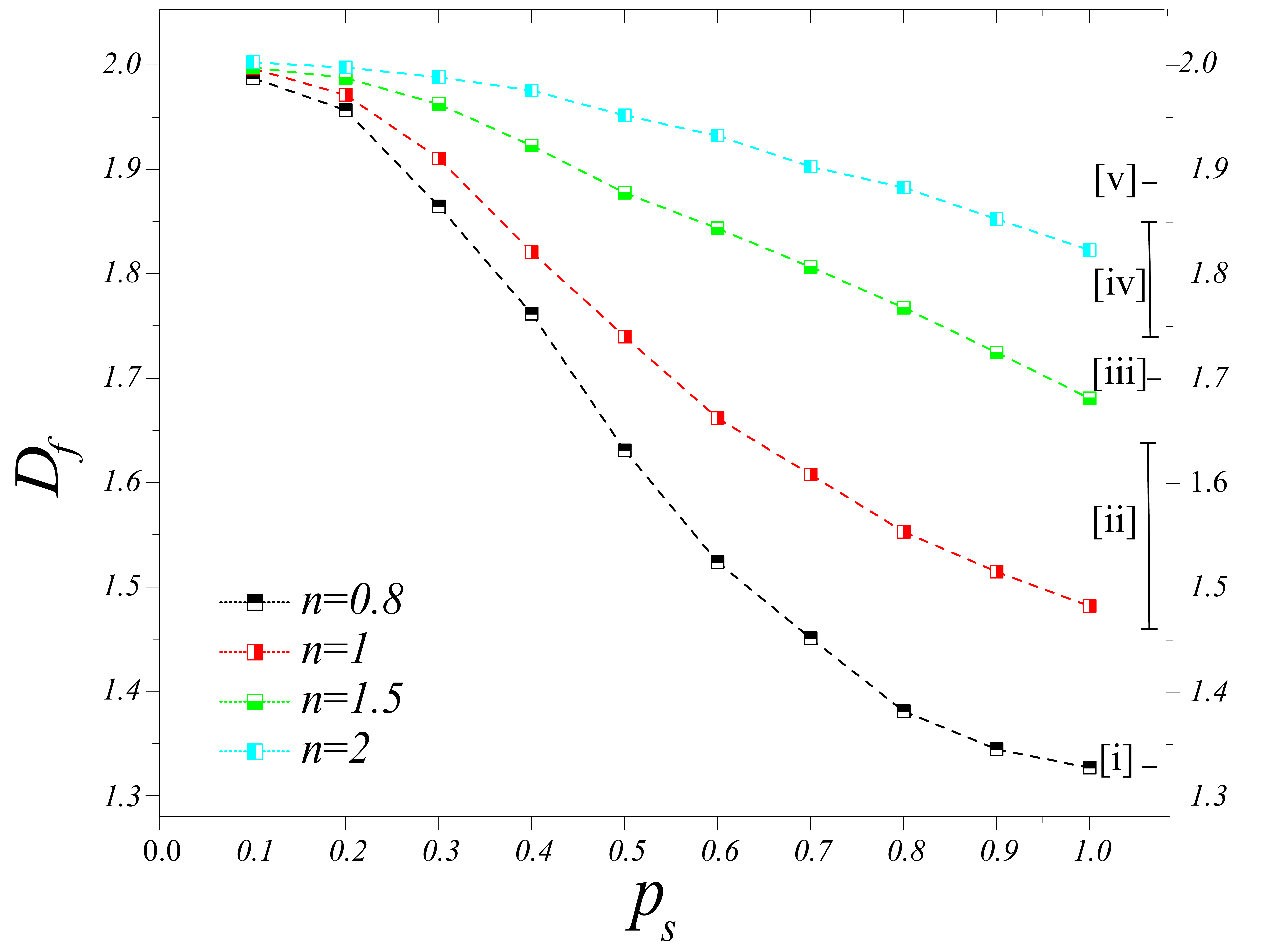}} 				
	\caption{\textbf{Interrelationship between malignancy, immortality and \textbf{oxygen} density.} Fractal dimension $D_f$ as an indication of malignancy for various tumors. Our model reproduces some of previously observed fractal dimensions: [i] $D_f\sim 1.338\pm 0.248$ \cite{smitha2015fractal}, [ii] 1.46 <$D_f$<1.64\cite{perez2015relationship}, [iv]  1.74<$D_f$< 1.85 \cite{pribic2015fractal}, [iii]  $D_f\sim 1.696 \pm 0.009$ and [v]  $D_f\sim 1.887 \pm 0.008$ \cite{buczko2015shape}.}
	\label{FIG2}
\end{figure}

A study of the variations of $D_f$ with $p_s$ and the density $n$ of the oxygen is useful to characterization of the tumor behavior. The computed  $D_f$ for various values of $p_s$ and oxygen densities are shown in Figure \ref{FIG2}.

Figure\ref{FIG2} presents explicitly the value of $D_f$ and the corresponding malignancy of tumor as a
result of both the internal feature and the external conditions.  For a fixed density 
$n$ of oxygen, the invasive behavior of tumor always increases with $p_s$, implying that, regardless of 
the environmental conditions, higher numbers of CSCs always lead to a more invasive behavior ---see Figure2
in the Supplementary Information (SI). This result contradicts the existing result on adverse effect of 
$p_s$ on the tumor's invasive behavior \cite{sottoriva2010cancer, enderling2013cancer}. On the other 
hand, the effect of the environmental stress on invasion is regulated by internal feature of cells, 
$p_s$. For $p_s=1$, the oxygen deprivation significantly increases the malignant behavior of tumors, 
while for $p_s=0$, the density of oxygen has negligible effect on tumor's invasive behavior. 

\section*{Relation to Superficial Spreading melanoma}
As presented here, our model explains a two dimensional tumor growth. Early stages of Superficial Spreading melanoma has a two dimensional structure which might be a good option to apply our findings to.  
Experiments indicate that there is no blood flow to the SSMs with thickness less than 0.9 mm \cite{srivastava1986neovascularization}. In addition, melanoma is, at least in its early
stages, an approximately two-dimensional (2D) phenomenon and a 2D model properly mimics its structure.
The malignant cells in the Superficial Spreading melanoma (SSM) stay within the original tissue - the epidermis - in an {\it in-situ} phase for a long time, which could be up to decades. Initially, the SSM grows horizontally on the skin surface, known as {\it radial growth}, with lesion indicated by a slowly-enlarging flat area of discolored skin. Then, part of the SSM becomes invasive, crossing the base membrane and entering the dermis, giving rise to a rapidly-growing nodular melanoma within the SSM that begins to proliferate more deeply within skin. 

\section*{Discussion}
The proposed model sheds new light on and provides new insight into the invasive behavior of tumors by deciphering the effect of both intrinsic and extrinsic features of cells.  It also
demonstrates that elimination of the oxygen in the previous models gives rise to such a 
relation. The fractal behavior that has been identified and then attributed to  the growth limited  to the perimeter, similar to surface growth\cite{bru1998super, bru2003universal}. Nevertheless, close
inspection of the proliferation activity in the perimeter in the proposed model reveals larger parts of 
the cells as proliferative cells ---see Figure1 of the Supplementary Information. As the model 
demonstrates, a single biological parameter, namely $p_s$, changes the cell's features and 
results collectively in various self-similar states with distinct fractal dimensions. Previous models,
which considered the CSCs \cite{sottoriva2010cancer, enderling2013cancer}, obtained an inverse relation 
between the number of the CSCs and invasion, but our model indicates increased malignancy to be 
proportional to larger numbers of the CSCs. 
Compared to experimental data\cite{castellon2012molecular} our model confirms   increasing of morphological irregularities (Gleason grade), but full consistency needs more biological details to be added to model. 

Tumors with low number of the CSCs that were proposed by the 
previous studies \cite{hermann2007distinct, enderling2013cancer} did not respond to 
oxygen  deprivation, as was expected \cite{hockel1996association,cristini2005morphologic}. Hence, tumors
that respond to  oxygen deprivation must have larger number of the CSCs.

In addition, models that do not consider the CSC evolution and endow the cells with unlimited 
proliferation capacity \cite{cristini2005morphologic,anderson2006tumor}, produce tumors corresponding to 
$p_s=1$. Such models consider the effect of oxygen and, as our model confirms, oxygen deprivation leads 
to higher irregularities. As $p_s$ decreases, the effect of  oxygen  vanishes. Thus, a 
lower number of the CSCs, which was proposed previously \cite{hermann2007distinct,enderling2013cancer}, 
does not conform to the experimentally well-established  oxygen  effects.  Our model, in addition to reproducing
such result, provides quantitative and comparable results to classify the irregularities that can be used
to study experimental results that have been reported the fractal dimensions. 

The conceptual results are applicable to the growth of 
	other solid tumors which display the mentioned behavior in response to oxygen tension and frequency of cancer stem cells. For example, in the case of the SSM in which the number of CSCs is not small \cite{girouard2011melanoma,quintana2008efficient}  oxygen  deprivation probably  increases tumor malignancy.  Contrary to the previous studies, the present model predicts invasion as the result 
of {\it both} the tumor {\it and} the microenvironment, demonstrating the effect of nutrient deprivation 
on the invasion. This implies that recent studies on such therapeutic approach
\cite{tang2016cystine,li2016dt} must consider carefully the side effects that, based on our model for
tumors with larger numbers of the CSCs, can increase tumor malignancy.

\section*{The model} 
\begin{figure}[h]
	\centerline{\includegraphics[width=1.0\linewidth]{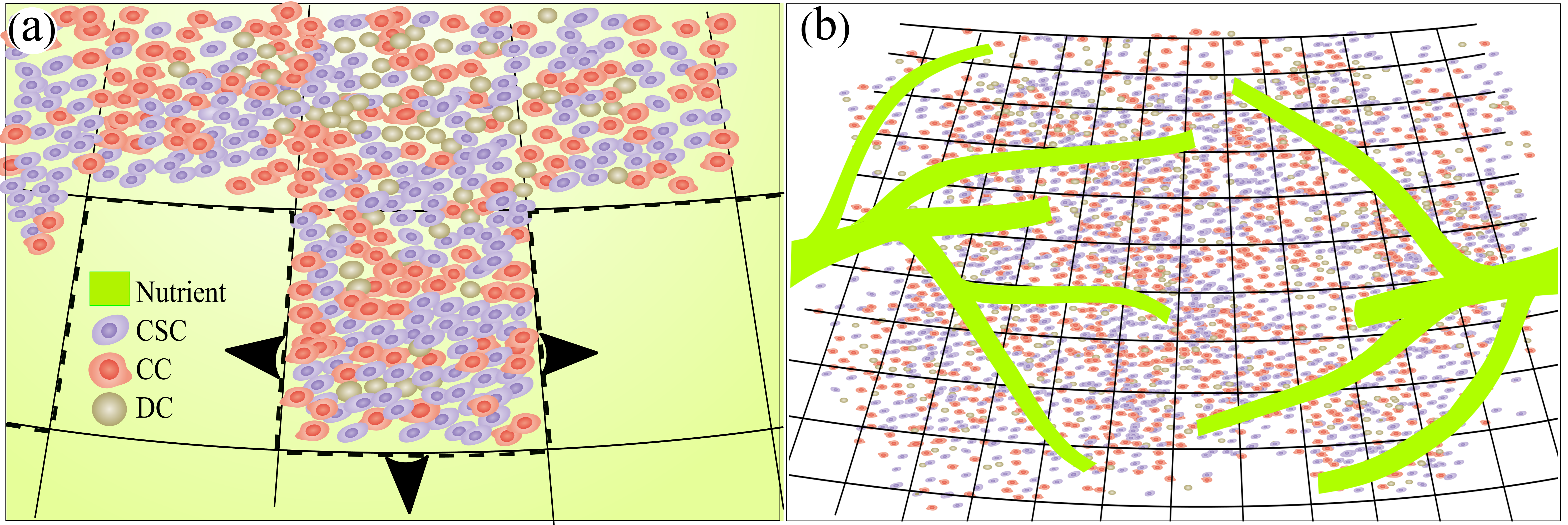}}
	\caption{\textbf{Schematic of the model.} (a) Various kinds of cells that are either proliferating or 
		dying. Nutrient density in the milieu is constant and after diffusing from the surrounding is consumed 
		by the cells. (b) An alternative mechanism for oxygen supply by the capillaries coming from the third
		dimension to feed the tumor at random sites. The results do not depend on the choice of the 
		initial/boundary conditions for the nutrient; see the SI.}
	\label{FIG3}
\end{figure}

Similar to many other natural systems, biological media fluctuate due to the intrinsic randomness of the 
individual events \cite{hilfinger2011separating}. Cells are involved in regulatory pathways that depend 
highly nonlinearly on the chemical species that are present in low copy numbers per cell 
\cite{berg2000fluctuations}, as a result of which other factors, such as the forces between cells, 
fluctuate significantly \cite{trepat2009physical}. Thus, statistical approaches are suitable for 
simulating cells' behavior. We consider the 2D lattice shown in 
Figure\ref{FIG3} in which each bond is 100 micrometer long, while each site has the capacity for 100 
cancer cells which typically have 10 $\mu$m diameter\cite{wang2011fiber}. The nutrient density is constant on the perimeter of a circle with a radius of 1 cm. It diffuses into the internal zones and is consumed by the living cells. In the 
Supplementary Information we present the results of various other initial/boundary conditions for 
the oxygen supply, including smaller and  larger radii of the circle, regular and random distribution of 
the  oxygen source, as well as its uniform distribution in the medium, and show that the
predictions of the model do not depend on the choice of the  oxygen  supply mechanism.  Though we considered two dimensional structure,  a 3D structure for oxygen supply system (vessels and capillaries), the results remain qualitatively the same. The model can, however, be  extended to 3D.\\

Keeping the  oxygen  density uniform in the milieu ---$0.15$ mol/ml \cite{anderson2005hybrid}--- a CSC is inserted at the center of medium that consumes the oxygen  and enhances its metabolic state. Although metabolic pathways are not fully understood, metabolic activity is a crucial factor in a cell's decision to either proliferate or die \cite{buchakjian2010engine}. In the former case a cell must increase its biomass and replicate its genome prior to division, in order to create two daughter cells. Thus, the cell must generate enough energy and acquire or synthesize biomolecules at a sufficient rate to meet the demands of proliferation \cite{jones2009tumor}. Given such biological facts, we choose metabolic state as the decisive factor for a cell's decision to proliferate, and define an internal energy $u_{\rm cell}$ for each cell as an 
indicator of its metabolic state. Physically, the cells acquire energy from the environment to accumulate
internal energy \cite{scalerandi2002inhibition}---the energy of the absorbed molecules--- which evolves 
according to the energy conservation law:
\begin{equation}
	\frac{\partial u_{\rm cell}}{\partial t}=\chi n(x,y,t)-\gamma u_{\rm cell}\;,
	\label{Eq1}
\end{equation}
where $n(x,y,t)$ is the oxygen density at position $(x,y)$ and time $t$, with $\chi$ and 
$\gamma$ being positive constants related to energy accumulation and consumption rate (for details about all constants and their values see Table 1 in SI). If a cell's energy reaches a threshold $u_p$, it will begin 
duplication. We set $u_p$, $\chi$ and $\gamma$ such that every cell in the appropriate situation will be 
in the duplication state after 15 hours \cite{haass2014real}, which is  about  the time 
that tumor cells need to reach the so-called cell checkpoints $eG_1$ (early $G_1$), $G_1$ and 
$eS$ in the cell cycle for division. $G_1$ is the primary point at which a cell must decide whether to 
divide. After it passes $G_1$ and enters the $S$ phase, the cell is committed to division 
\cite{haass2014real} (other checkpoints, such as $G_2$ at which the cell is mostly concerned with the 
condition of its DNA, still remain to be completed in the next step). As we show below, Eq. (\ref{Eq1}) 
together with the limits imposed reproduces cell plasticity and various proliferation activities under a 
variety of external conditions \cite{meacham2013tumour} that were reported recently \cite{haass2014real}.
Time is measured in units of 10 minutes. 

\begin{figure}[h]
	\centerline{\includegraphics[width=0.6\linewidth]{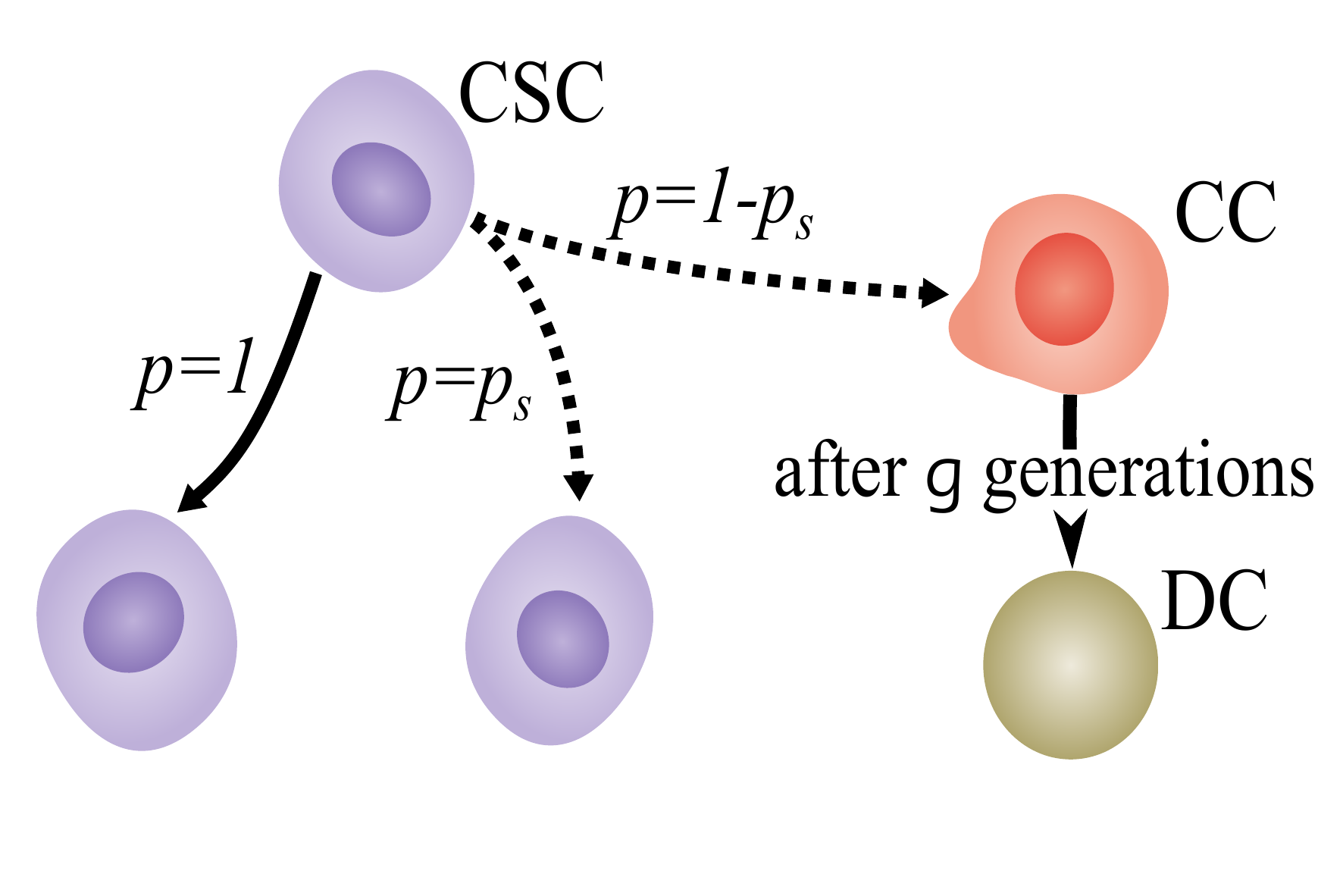}}
	\caption{\textbf{Division of the cells.} During division each CSC creates another CSC. The other 
		daughter cell would either be a CSC with probability $p_s$ or a CC with probability $1-p_s$. 
		Each CC creates two CCs during duplication, if it is capable of division.  The CSCs 
		can continue the division process for a long time, whereas each CC loses its ability for duplicating 
		after $g$ divisions, and dies. Clearly, the first CC daughters could duplicate $g-1$ times, where we set 
		$g=5$ \cite{sottoriva2010cancer}.}
	\label{FIG4}
\end{figure}

The evolution of the internal energy $u_{\rm cell}$ of the cells depends on the local 
density of oxygen through a set of coupled differential equations, and if enough oxygen exists at the 
position of the first CSC, $u_{\rm cell}$ increases to $u_p$ and the first CSC duplicates into two 
daughter cells. This relation between oxygen density, cell metabolic state and its duplication dynamics 
ensures the apparent role of the oxygen density in the tumor evolution.  One may consider various 
scenarios for quantitative studies of the CSC proliferation 
\cite{shahriyari2013symmetric,dhawan2014tumour,tomasetti2010role,cao2013modeling}, but the probability of
distinct kinds of divisions has yet to be assessed experimentally. Besides, some other 
studies \cite{yoo2008cancer} have proposed the cells' self-renewal ability as the prerequisite for tumor 
maintenance. Thus, we choose the simplest biologically-correct model that has the ability to generate 
the entire possible range of the CSC population percentage, from zero up to the values produced by the 
various mathematical \cite{shahriyari2013symmetric,dhawan2014tumour,tomasetti2010role,cao2013modeling} 
and biological models \cite{yoo2008cancer}.  In this model, during duplication  of each 
CSC  one daughter cell is assumed to be CSC and the second one is either a CSC with probability $p_s$ ---the probability of symmetric duplication of the CSCs--- or a cancerous cell (CC) with probability 
$(1-p_s)$ ---see Figure\ref{FIG4}.  Each CC duplicates into two CCs if it is allowed to duplicate 
\cite{sottoriva2010cancer}. Such a probabilistic approach is motivated by a fact stated earlier, that 
according to the classical CSC hypothesis, among all cancerous cells, only ``a few'' act as stem cells, 
whereas some studies\cite{quintana2008efficient, gedye2016cancer} have reported that the population of CSCs can be relatively high, which is why we take the population of the CSCs (the probability $p_s$) as 
a parameter of our model. For $p_s=1$ the model reduces to the stochastic model of tumor development 
\cite{Nowell1976}. Every CSC continues such a division for an unlimited frequency, but the CC can have 
only limited generations of duplication \cite{hayflick1961serial}, which  we set it to be $g=5$ 
\cite{sottoriva2010cancer,reya2001stem} after which it will die and produce dead cells (DCs) ---see Figure\ref{FIG4}. As the cells undergo apoptosis, they are recognized and removed from the body by phagocytes. 
Thus, we assume that the dead cells remain inactive in the medium, but even if we eliminate them after 
death, the main results  remain the same; see the Figure S15 at SI.\\

We define the  density  of cells of type $i$ at location $(x,y)$ at time $t$ by,
\begin{equation}
	C_i(x,y,t)=\frac{\text{number of cells at}\;(x,y,t)}{\text{capacity of each site}}\;,
	\label{Eq2}
\end{equation}
with $i\equiv$ CSCs, CCs, and DCs. Equation (\ref{Eq2}) is also valid for the total 
density of cells, $C_t=C_{\rm CSC}+C_{\rm CC}+C_{\rm DC}$. Recall also that the capacity of each site 
is 100 cells \cite{wang2011fiber}. The  density of the CCs is denoted by $
C_{\rm CC}(x,y,t;j)$ in which $j$ indicates their generation that varies from 1 to $g$ (after $g$
generations they produce the DCs). Healthy tissues contain healthy cells in which the distribution of the
nutrients is in a steady state. We eliminate the healthy cells for all the tumors, as our results are 
based on comparison with and differences of tumors' behavior that are the most important part of our 
study.\\

Local  density gradients drive  the random motion of the cells 
\cite{ambrosi2002closure}. Thus, one has,
\begin{equation}
	\frac{\partial C(x,y,t)}{\partial t}=D\nabla^2C(x,y,t)\;,
	\label{Eq3}
\end{equation}
where $D$ is the  diffusion coefficient. Equation (\ref{Eq3}) is applicable to the 
various kinds of cells, for which \cite{anderson2005hybrid,bray2001cell} $D\approx 10^{-10}$cm$^2$/s. 
Population growth of biological groups depends on the species ability for proliferation and the 
environmental limitations. One important environmental limit is contact inhibition
of cell division \cite{martz1972role}, i.e., if after the energy rises to $u_p$ the cells will duplicate,
if there is space; otherwise, they will stay quiescent  until they find space for 
duplication\cite{montel2011stress}. Thus, proliferation at each site depends on the number of cells 
that can duplicate, and the effect of competition for space between all types of cells. The evolution of 
the CSCs  that qualifies for the duplication metabolic threshold $u_{p}$,  is expressed 
by a diffusion-reaction equation,
\begin{eqnarray}
	\frac{\partial C_{\rm CSC}(x,y,t)}{\partial t}&=& D\nabla^2C_{\rm CSC}(x,y,t) \\
	& &+R_mp_{s}C_{\rm CSC}(x,y,t)[1-C_t(x,y,t)]\nonumber,
	\label{Eq4}
\end{eqnarray}
where $R_m$ is the rate of passing the $S$, $G_2$ and $M$ phases in the cell cycle, which is fixed as a
cell that has enough internal energy (has passed the aforementioned $eG_1$, $G_1$ and $eS$ phases) will 
duplicate in 5 hours \cite{haass2014real}, if there were no other cells. The last term
on the right side of Eq. (\ref{Eq4}) that includes the term $[1-C_t(x,y,t)]$ captures the effect of 
contact inhibition of proliferation in which $C_t(x,y,t)$ is the total density of all cells at 
$(x,y,t)$ . The entire cell cycle takes 20 h. The evolution of the $j$th generation of the CCs is 
governed by
\begin{eqnarray}
	\frac{\partial C_{\rm CC}(x,y,t;j)}{\partial t}= D\nabla^2 C_{\rm CC}(x,y,t;j)\nonumber\\
	+ \delta_{1j}R_m[1-p_s][1-C_t(x,y,t)] 
	\nonumber \\+ (1-\delta_{1j})R_mC_{\rm CC}(x,y,t;j-1)[1-C_t(x,y,t)]
	\nonumber \\-    (1-\delta_{jg})  R_mC_{\rm CC}(x,y,t;j)[1-C_t(x,y,t)]
	\nonumber \\- \delta_{jg}R_aC_{\rm CC}(x,y,t;j)\;,
	\label{Eq5}
\end{eqnarray}
where $\delta_{ij}$ denotes the Kronecker delta, i.e., $\delta_{ij}=1$ for $i=j$ and $0$ otherwise, with $1\leq i,j\leq g$. The first term on the right side of Eq. (\ref{Eq5}) represents diffusion of the cells due to the local concentration gradient 
\cite{ambrosi2002closure,anderson2005hybrid}; the second is the creation of the first generation of the 
CCS due to asymmetric duplication of the CSCs \cite{sottoriva2010cancer}, while the third term represents
the creation of the $j$th generation (for $j\ne 1$) of the CCs from duplication of the prior generation. 
The concentration of the CCs decreases due to duplication and creation of the next generation, which the 
$4$th terms accounts for, while the last term accounts for the death of the final ($g$th) generation of 
the CCs. $R_a$ is the rate of apoptosis ---the process of programmed cell death--- and is fixed as the 
$g$th generation has a halflife equal to 1 day. Finally, the evolution of the oxygen density in the 
presence of the cells is governed by
\begin{eqnarray}
	\frac{\partial n(x,y,t)}{\partial t}&=&\beta\nabla^2 n(x,y,t) \\&&-\alpha[C_{\rm CSC}(x,y,t)+\sum^g_{j=1}
	C_{\rm CC}(x,y,t;j)]\nonumber,
	\label{Eq6}
\end{eqnarray}
with $\alpha$ being proportional to oxygen consumption rate by the cells, which is the same  for both 
the CCs and cancerous stem cells. We varied the rates of oxygen consumption for every kind of cells, but 
the essential results remained the same; see the SI. $\alpha$ was fixed by setting the 
reported value for oxygen consumption \cite{casciari1992variations,anderson2005hybrid} to be $6.65\times 
10^{-17}$ mol cell$^{-1}$s$^{-1}$. $\beta$ is the diffusion coefficient of oxygen in 
the medium, which we fixed it based on the calculations at room temperature, $10^{-5}$ cm$^2$/s. We 
present in the SI the results for other values of $\beta$. For distances more than 1 cm from the medium's
center the oxygen density is constant (see the SI for the results for larger and smaller distances, as 
well as other ways of supplying the oxygen), and is equal to 0.15 mol/ml \cite{anderson2005hybrid}. For
simplicity, in all the calculations we normalize $n$ to 1. From outside of the aforementioned circle,
oxygen penetrates into the central area. Given the assumptions, the cells are active elastic species,
consuming oxygen and proliferating. 

As we show in the SI, other boundary conditions do not change the essential results. In addition, (i) we 
also varied both the proliferation activity and oxygen consumption rate for various kinds of cells, but
the results remained qualitatively the same. (ii) The CSCs and CCs are assumed to have 
equal oxygen consumption rates, but when we changed them for every kind of cell, the 
results were qualitatively the same. (iii) The CSCs and CCs are assumed to have the same internal energy
threshold $u_p$ for duplication, and equal rates of crossing the $S$, $G_2$ and $M$ phases in the cell 
cycle. But changing the proliferation activity of the cells did not change our main results. Let us also 
emphasize that our model is not the same as the classical models of diffusion-limited aggregation 
\cite{gerlee2010diffusion}, as such model did not deal with the effect of reaction and consumption.

\section*{Acknowledgements}

A.A.S. would like to acknowledge supports from the Alexander von Humboldt Foundation, and partial 
financial supports from the research council of the University of Tehran. We also acknowledge the High 
Performance Computing center of the University of Tehran in its Department of Physics, where most of 
computations were carried out.

\bibliography{draft}
\end{document}